\documentclass[10pt,notitlepage]{article}
\usepackage{amssymb}
\usepackage{amsmath}
\usepackage{graphicx}

\newcommand{\rd}{\mbox{$d$}}
\newcommand{\half}{\mbox{$\textstyle \frac{1}{2}$}}
\newcommand{\quat}{\mbox{$\textstyle \frac{1}{4}$}}

\title{Relativistic state reduction dynamics}
\author{
Daniel~J.~Bedingham
\footnote{{\it Blackett Laboratory, Imperial College, London SW7 2BZ, UK.}}
\footnote{email: d.bedingham@imperial.ac.uk}
}
\date{\today}

\begin{document}
\maketitle

\begin{abstract}
A mechanism describing state reduction dynamics in relativistic quantum field theory is outlined. The 
mechanism involves nonlinear stochastic modifications to the standard description of unitary state 
evolution and the introduction of a relativistic field in which a quantized degree of freedom is 
associated to each point in spacetime. The purpose of this field is to mediate in the interaction 
between classical stochastic influences and conventional quantum fields.
The equations of motion are Lorentz covariant, frame independent, and do not result in divergent 
behavior. It is shown that the mathematical framework permits the specification of unambiguous local 
properties providing a connection between the model and evidence of real world phenomena. 
The collapse process is demonstrated for an idealized example.

\vspace{10pt}
\noindent
PACS numbers: 03.65.Ta, 11.10.-z, 02.50.Ey.

\end{abstract}

\section{Introduction}

The pragmatically applied rules of quantum mechanics involve two distinct laws of 
evolution for the state of the system: these are the Schr\"odinger equation and quantum state reduction. 
A long-standing problem is how to make sense of this since there is no underlying theory stating when one or 
the other of these laws is to be used. Instead it is left to a judgment whereby state reduction is associated with 
the fuzzy concept of measurement.

Based on the premise that quantum state reduction should be taken seriously as a genuine physical process, 
collapse models \cite{ghir3,ghir2} are an attempt to resolve this situation by suggesting a composite 
dynamics incorporating state reduction events or collapses and unitary state evolution (for general reviews see \cite{Bass, Pear2}). 
The idea is that the Schr\"odinger equation should be viewed as an approximation to this more general dynamics
valid when collapse effects are negligible. Conversely, collapse effects should be seen to dominate in situations 
where state reduction is an appropriate description.

The most familiar model of this type is that of Ghirardi, Rimini, and Weber (GRW) \cite{ghir3} describing a system 
of nonrelativistic quantum particles. The essential idea of GRW is that the state of each particle, as a matter of 
physical law, occasionally (but very infrequently) undergoes a random collapse to a state localized in position space. 
From this law of collapses it follows that quantum wavelike behavior becomes increasingly unstable for systems of increasing 
size \cite{ghir3}. Collapse models thus offer a mathematical framework capable of unifying quantum and classical 
domains.

The nonrelativistic Continuous Spontaneous Localization (CSL) model \cite{ghir2} is an improvement on 
GRW since it preserves the symmetries of systems of identical particles. The formulation of CSL, in terms of a
stochastic differential equation, invites a straightforward generalization to relativistic quantum field theory (QFT), 
but it is well known that this results in physically unacceptable divergent behavior \cite{pear3,pearGhir}.
Here we shall address the question of how these infinities can be avoided.

The source of divergences, as with other infinite behavior in QFT, can be traced 
to point interactions between quantum field operators in the dynamical equations for the state vector. 
However, in the case of relativistic collapse models, attempts to renormalize with the inclusion of subtractive counter 
terms fail. This way of viewing the problem of infinities suggests that a solution 
could be to smear out the point interactions. The same idea was considered by Nicrosini and Rimini \cite{Nicr} 
although their implementation requires the unsatisfactory inclusion of a locally preferred frame. 
In this article we propose the use of a novel relativistic field responsible for mediating the collapse 
process which enables us to fulfill the aim of smearing the interactions whilst preserving Lorentz covariance and 
frame independence. This forms the basis of a relativistic collapse mechanism which naturally resembles 
CSL, describing state reductions in a smeared number density eigenbasis.

The structure of the article is as follows. We begin the presentation of our model in section \ref{pf} by stating the 
properties of the mediating field (which we subsequently refer to as the pointer field). In section \ref{sd} we define 
the dynamical equations of motion following the outline for relativistic collapse models given by Pearle \cite{pear3}. 
We discuss the local properties of the model in section \ref{lb} and the form of the smeared interactions in 
section \ref{so}. 

In section \ref{cp} we describe the collapse process in detail for an idealized example. We estimate
a collapse timescale and demonstrate that the dynamics reproduce the Born rule. In section \ref{ep} we show that the 
model does not exhibit physically unacceptable divergent behavior by considering how the energy of the system is 
influenced by the equations of motion. We end with a numerical demonstration of the collapse process and a short discussion.

\section{Pointer field}
\label{pf}

Consider a field in which a quantized degree of freedom is associated to each point in spacetime. 
This is to be contrasted with a standard quantum field whose modes describe the field configuration on
a time slice or spacelike hypersurface.
We define creation and annihilation operators $a(x)$ and $a^{\dagger}(x)$ with commutation relations 
\begin{align}
[a(x),a^{\dagger}(x')] = \delta^4(x-x') \quad ; \quad  [a(x),a(x')] = 0,
\end{align}
and specify a normalized ground state with the property that $a(x)|0\rangle=0$. 

First excited states are 
given by
\begin{align}
|h\rangle = \int d\omega_x h(x)a^{\dagger}(x)|0\rangle,
\end{align}
where $\rd\omega_x$ denotes the integration measure over spacetime volume and 
$h$ is some complex $L^2$-function on spacetime. 
Higher exited states can be constructed by repeated application of the creation operator in this way and we define 
our state space to accommodate addition of states (enabling field superpositions).
To ensure that the field transforms appropriately when specified in different coordinate frames we supply it with the 
transformation property
\begin{align}
U_{\Lambda,b} a(x) U^{-1}_{\Lambda,b} = a(\Lambda x+b),
\end{align}
for a Lorentz coordinate transformation $\Lambda$, spacetime translation $b$, and unitary representation of the 
Poincar\'e group $U$. 

We refer to this general construction as the pointer field since its role in our model is to make a record of 
the state of a conventional quantum field with which it interacts. An equivalent construction used for a different purpose 
can be found in references \cite{CQC1,CQC2}. The pointer field's degrees of freedom describe a field configuration over
the whole of spacetime.

\begin{figure}[t]
\begin{center}
\includegraphics[width=8cm]{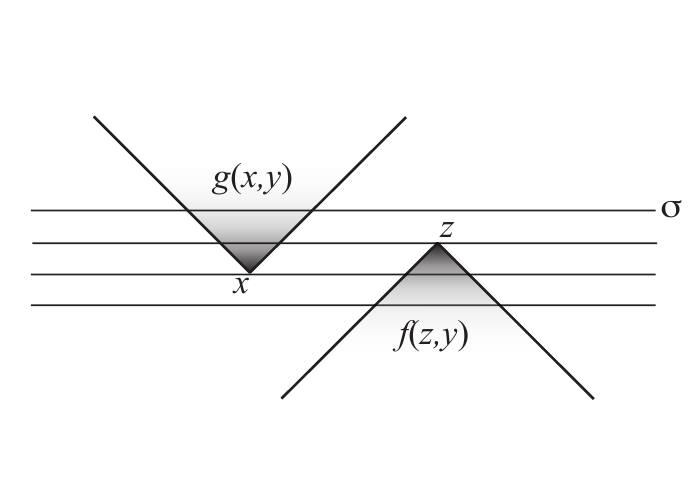}
\caption{
{A representation of the domains of functions $f$ and $g$ in spacetime. 
The points $x$ and $z$ are spacelike separated and $\sigma$ denotes a spacelike hypersurface belonging 
to some spacetime foliation. 
}}
\end{center}
\end{figure}

The number density operator is given by $n(x) = a^{\dagger}(x)a(x)$. We use this to construct new operators which are 
smeared over spacetime
\begin{align}
N(x) = \int \rd \omega_y f(x,y) n(y).
\end{align}
Here $f(x,y)$ is some invariant function on 
spacetime which is only nonzero for $y$ in the past cone of $x$ (see figure 1). Similarly we define
\begin{align}
A(x) = \int \rd\omega_y g(x,y) \left[a(y)+a^{\dagger}(y)\right]
\end{align}
where $g(x,y)$ is some invariant function which is only nonzero for $y$ in the future cone of
$x$ (figure 1). Proposals for the functions $f$ and $g$ will be specified in section \ref{so}. Note that
\begin{align}
[N(x),N(x')]=0 \quad {\rm and}\quad [A(x),A(x')] = 0 \quad \forall\;\; x,x'.
\label{0com}
\end{align}
However, 
\begin{align}
[N(x),A(x')]= \int \rd\omega_y f(x,y)g(x',y)\left[a^{\dagger}(y) - a(y)\right],
\label{commie}
\end{align}
where the right hand side is only nonzero when $x$ is in the future cone of $x'$. This entails that 
$[N(x),A(x')]=0$ if $x$ and $x'$ are spacelike separated. 

We regard the pointer field as a new and fundamental component of our model (rather than as some effective construction
representing the effects of standard quantum fields). 

\section{State dynamics}
\label{sd}

For a relativistic collapse model we require a covariant description of how the state changes as we 
advance through spacetime. The dynamics should involve a classical stochastic input to capture 
the random character of quantum state reduction and we expect our equations to be nonlinear reflecting a feedback
from the state vector to the probability of an outcome.

\subsection{Tomonaga picture}

Consider the orthodox dynamics of a conventional relativistic quantum field. In order to form a covariant description of the 
evolving state of the field we use the Tomonaga picture: A state $|\Phi(\sigma)\rangle$ is assigned to any given 
spacelike hypersurface $\sigma$. As we advance the surface $\sigma$ to a new surface $\sigma'$ which differs from $\sigma$ only 
at the point $x$ such that $\sigma$ and $\sigma'$ enclose an incremental spacetime volume $d\omega_x$, the change of
state is given by the Tomonaga equation
\begin{align}
\rd_x |\Phi(\sigma)\rangle = 
-i H_{\rm int}(x) \rd \omega_x|\Phi(\sigma)\rangle,
\label{TOM}
\end{align}
where $H_{\rm int}(x)$ represents the interaction Hamiltonian. Any interaction terms must be Lorentz scalars to give the equation 
covariant form and must commute at spacelike separation to reflect the fact that there is no temporal ordering of spacelike 
separated points (i.e.~no preferred frame). Aharonov and Albert argue in reference \cite{aa} that, for a covariant description
of state collapse, the state must take the form of a functional on the set of spacelike hypersurfaces as in this picture.  

For our model we consider a state which describes both a quantum field and pointer field. Given the commutation 
relations~(\ref{0com}) and (\ref{commie}) and given the above constraints on $H_{\rm int}$ we may use the Tomonaga 
picture to describe the evolving state where $H_{\rm int}$ is constructed from terms involving $N(x)$ and $A(x)$ 
(along with quantum field operators). This allows us to describe state evolution involving interactions between the 
quantum field and pointer field. We remark that whereas the quantum field state describes the quantum field on some 
given hypersurface, the pointer field state describes the pointer field over the whole of spacetime. 
The pointer field state nevertheless depends on the given hypersurface since this demarcates a boundary of past 
interactions with the quantum field.

In the Tomonaga picture we are required to think of state evolution with regards to an ordered sequence of 
spacelike hypersurfaces. The relationship between different spacelike hypersurfaces in spacetime can be 
classified by a partial ordering structure. Consider two surfaces $\sigma_1$ and $\sigma_2$. If no point 
in $\sigma_1$ is to the causal future of any point in $\sigma_2$ then we can say that $\sigma_1 \prec \sigma_2$. 
(We will also use the notation $\sigma \prec x$ and $x\prec\sigma$ to denote that the point $x$ is not to the past
and not to the future of $\sigma$ respectively.) The partial order relation $\prec$ is
\begin{align}
 {\rm reflexive\;\; : \;\;} & \sigma \prec \sigma,
\nonumber\\
 {\rm antisymmetric\;\; : \;\;} & (\sigma_1 \prec \sigma_2) \land (\sigma_2 \prec \sigma_1) \Rightarrow \sigma_1 = \sigma_2,
\nonumber\\
 {\rm transitive\;\; : \;\; } & \sigma_1 \prec \sigma_2 \prec \sigma_3 \Rightarrow \sigma_1 \prec \sigma_3.
\end{align}
A foliation of spacetime is any maximally ordered chain of surfaces. In a model with no preferred frame the 
foliation should have no physical significance---it should be considered to be analogous to a choice of gauge.

\subsection{Stochastic processes}

In order to understand the disclosure of stochastic 
information in the context of hypersurfaces advancing through a 
foliation of spacetime we require an appropriately structured probability space.
We specify our probability space by
$(\Omega, {\cal F}, \mathbb{Q})$ along with a filtration $\{{\cal F}_{\sigma}\}$ of ${\cal F}$,
defined to be a family of sigma-algebras ${\cal F}_{\sigma}\subset {\cal F}$ such that 
\begin{align}
\sigma_1 \prec \sigma_2 \Rightarrow  {\cal F}_{\sigma_1} \subset {\cal F}_{\sigma_2}.
\end{align}
The partially ordered set structure of the spacelike hypersurfaces is thus induced on the subset structure 
of the filtration. The subsets $F$ of $\Omega$ belonging to ${\cal F}$ are the events of our probability
space (e.g. $F=\{\text{the state assigned to $\sigma$ is }|0\rangle\}$). 
We interpret $\mathbb{Q}(F)$ as the probability that the event $F$ occurs. 
The construction of a filtration on the probability space allows us to formalize the notion that the consequences 
of the outcome of chance (an element $\omega$ of $\Omega$) are not necessarily revealed at once, but rather may emerge 
sequentially as the system evolves. This is achieved using the concept of conditional expectation with respect to 
${\cal F}_{\sigma}$, having the intuitive meaning of conditioning with respect to information about the set 
of events belonging to ${\cal F}_{\sigma}$. Stochastic processes in this context are random variables indexed by $\sigma$.

We describe the classical stochastic input of our model in terms of a noise field on spacetime.
By comparison with standard Brownian motion we can define a Brownian motion field in terms of infinitesimal 
increments $\rd W_{x}$ specified at each spacetime point with properties
\begin{align}
\mathbb{E}^{\mathbb{Q}}[\rd W_{x}] = 0 \quad \text{and} \quad \rd W_{x}\rd W_{x'} = \delta_{x,x'} \rd\omega_x,
\end{align}
where $\mathbb{E}^{\mathbb{Q}}[\; \cdot \;]$ denotes $\mathbb{Q}$-expectation.
We assume that the filtration $\{{\cal F}_{\sigma}\}$ is generated by our Brownian motion field such that for
any $x$ where $\neg(\sigma \prec x)$,  $\rd W_x$ is ${\cal F}_{\sigma}$-measurable. We can define a Brownian motion
process $W_{\sigma}$ such that $W_{\sigma'}-W_{\sigma} = \int_{\sigma}^{\sigma'}dW_x$ for any $\sigma\prec \sigma'$.

\subsection{Implicit equation of motion}

Other than spacetime, the structure of our model involves three spaces: (i) the space $\Sigma$ of all possible 
spacelike hypersurfaces $\sigma$ in spacetime; (ii) a probability space $(\Omega, {\cal F}, \mathbb{Q})$ in which 
all $\rd W_x$ are specified; and (iii) a Hilbert space ${\cal H}$ which describes the degrees of freedom of our 
universe (including matter fields, gauge fields, and the pointer field). The model describes a joint map from 
$\Sigma$ and $\Omega$ to 
${\cal H}$
\begin{align}
\Phi  : {}& \{\Sigma, \Omega\} \rightarrow {\cal H}, \nonumber \\
& \{\sigma, \omega\} \mapsto |\Phi(\sigma, \omega)\rangle.
\end{align}
Given an initial condition for the state we can define this map in terms of state evolution by the stochastic differential equation
\begin{align}
\rd_x |\Phi(\sigma)\rangle = \left\{
-i J(x) A(x) \rd \omega_x - \half \lambda^2 N^2(x) \rd \omega_x +\lambda N(x) \rd W_x
\right\}|\Phi(\sigma)\rangle.
\label{SE}
\end{align}
We also specify a change of probability measure
\begin{align}
\mathbb{E}^{\mathbb{P}}[\; \cdot \;|{\cal F}_{\sigma}] = 
\frac{\mathbb{E}^{\mathbb{Q}}[\; \cdot \; \langle\Phi(\sigma_f)|\Phi(\sigma_f)\rangle |{\cal F}_{\sigma} ]}
{\mathbb{E}^{\mathbb{Q}}[\langle\Phi(\sigma_f)|\Phi(\sigma_f)\rangle |{\cal F}_{\sigma}]},
\label{PROB}
\end{align}
which relates the defining probability measure $\mathbb{Q}$, under which all Brownian increments $d W_x$
are independent, to the physical probability measure $\mathbb{P}$, under which stochastic probabilities
of evolved states agree with quantum predictions (see below). The surface $\sigma_f$ should be entirely to the future
of any regions of interest but is otherwise arbitrary owing to the fact that 
$\langle\Phi(\sigma)|\Phi(\sigma)\rangle$ is a $\mathbb{Q}$-martingale:
\begin{align}
\mathbb{E}^{\mathbb{Q}}[\langle\Phi(\sigma')|\Phi(\sigma')\rangle|{\cal F}_{\sigma}] 
=\langle\Phi(\sigma)|\Phi(\sigma)\rangle,
\end{align}
for $\sigma\prec\sigma'$ (the tower rule can then be used to show that $\mathbb{P}$-expectations for different 
$\sigma_f$ are equivalent). 

The stochastic coupling parameter $\lambda$ is a constant which relates to the rate at which the 
collapse process occurs and the Lorentz invariant operator $J(x)$ is a scalar current operator 
representing the matter density of a quantum field. (For example, we might choose $J(x) = \bar{\psi}(x)\psi(x)$ 
for a Dirac field $\psi(x)$.) We will refer to $J(x)$ as the matter density operator for the quantum field.
We have omitted any interactions between different quantum fields in equation (\ref{SE}), however, these can 
easily be added.

Equation (\ref{SE}) is a stochastic extension of the Tomonaga formulation of quantum state 
evolution. By setting $\lambda = 0$ we recover the Tomonaga equation in differential form (\ref{TOM}).
Provided that $J(x)$ commutes with $J(x')$ for spacelike separated $x$ and $x'$ then all terms in the 
evolution equation (\ref{SE}) commute at spacelike separation. (This is indeed the case for the example 
$J(x) = \bar{\psi}(x)\psi(x)$ where $\{\psi(x),\bar{\psi}(x')\}=\{\psi(x),\psi(x')\}=0$ for spacelike 
separated $x$ and $x'$.) This fact ensures that the specific foliation used has no physical consequences 
since given a fixed initial state and a complete realized set of stochastic information $\{dW_x\}$, for 
any two foliations which share a common leaf, the assigned state on that leaf is unique. Equation (\ref{PROB}) 
also shows that the physical probability density of a given realized set of stochastic information 
$\{dW_x |\sigma \prec x \prec \sigma_f \}$ conditional on ${\cal F}_{\sigma}$ depends only on the covariantly 
defined and foliation independent state norm assigned to $\sigma_f$. This in turn is determined from the state 
at $\sigma$ using the same realized stochastic information $\{dW_x |\sigma \prec x \prec \sigma_f\}$. 

Since the physical probability of obtaining a given final state 
depends on the final state itself we refer to this formulation as implicit. 
Equations (\ref{SE}) and (\ref{PROB}) completely specify the dynamics of the model in a covariant and frame 
independent manner.

\subsection{Collapse mechanism outline}
\label{CMO}

Consider the pointer field initially in its ground state. As the state evolves according to equation
(\ref{SE}), the interaction described by the term $J(x)A(x)$
leads to an excitation of the pointer field only in the future cone at $x$. We assume for now that the smearing density 
$g(x,y)$ is fairly well localized about $x$ in some sense. If $J(x)$ is significant at $x$ then the effect can
be thought of as analogous to that of a particle passing through a cloud chamber where a record of the track is formed 
and left behind. 

By contrast, the operator $N(x)$ acts on the pointer field state only in the past light cone of the point $x$,
registering the track made by $J$. 
The effect of the last two terms on the right side of equation (\ref{SE}) can be understood by considering an 
incremental stage in the state evolution. We can write 
\begin{align}
(1+\Delta_x) |\Phi(\sigma)\rangle &\sim \left\{1- \half \lambda^2 N^2(x) \Delta \omega_x +\lambda N(x) \Delta W_x
\right\}|\Phi(\sigma)\rangle \nonumber\\
&\sim \exp\left\{- \lambda^2 \left[N(x)-\frac{1}{2\lambda}\frac{\Delta W_x}{\Delta\omega_x}\right]^2 \Delta\omega_x + \quat
\right\}|\Phi(\sigma)\rangle.
\label{hercoll}
\end{align}
Heuristically we see that the state is acted on by a Gaussian positive valued operator which is centered 
about a point determined by the random choice of $\Delta W_x$. This has the effect of diminishing the quantum 
amplitude of all $N(x)$-eigenstates with respect to this central value. The probability rule (\ref{PROB}) is designed to 
ensure that the location of this projection is more likely where the quantum amplitude is greatest and in so doing
reproduce the Born rule (we will examine this in detail in section \ref{cp}). As the state evolves it is impelled by 
these projections toward an $N(x)$-eigenstate.
Reductions occur to a smeared number density eigenbasis as with the CSL model. This model can therefore be 
seen as a natural relativistic extension of CSL.

It is important that the excitations made to the pointer field influence 
the result of acting with the operator $N(x)$ at $x$. Equation~(\ref{commie}) indicates that this is the case 
provided that the excitation involving $A(y)$ occurs for $y$ in the past cone of $x$.

Since interactions between the quantum field and the pointer field result in entanglement between different quantum matter 
densities and different pointer field states, the collapse of a superposition of pointer field states will induce a collapse
of quantum field states. We will consider specific examples of this process in later sections.

\subsection{Nonlocality}

The probability rule (\ref{PROB}) is responsible for nonlocal correlations in this model. Consider a state which describes 
two spacelike separated subsystems of an entangled global system (such as in an EPR-type experiment). Suppose that each of these
subsystems undergoes a collapse (such as that involved in a spin measurement). The probability rule ensures not only that the outcomes of
the collapse processes for each subsystem occur individually with the correct quantum probabilities but also that the 
joint probabilities for outcomes in the two subsystems satisfy quantum predictions. In this case we find that the $dW_x$s
are correlated over spacelike separation in the physical probability measure. The $\mathbb{Q}$-Brownian motion field behaves as
a nonlocal hidden variable in the theory. This is analyzed in detail in reference \cite{ME}.

\subsection{Explicit equation of motion}

We can define a Brownian motion field under the $\mathbb{P}$-measure such that
\begin{align}
\mathbb{E}^{\mathbb{P}}[\rd B_{x}] = 0 \quad \text{and} \quad \rd B_{x}\rd B_{x'} = \delta_{x,x'} \rd\omega_x.
\label{BMotion}
\end{align}
Given a specific foliation of spacetime we can relate this to the $\mathbb{Q}$-Brownian motion field by defining
\begin{align} 
\rd B_x = \rd W_x -2\lambda\langle N(x) \rangle_{\sigma}\rd \omega_x,
\label{Pnoise}
\end{align}
where we have used the notation
\begin{align}
\langle \; \cdot \; \rangle_{\sigma} = \frac{\langle\Phi(\sigma) |\;\cdot\; |\Phi(\sigma)\rangle}{\langle\Phi(\sigma) |\Phi(\sigma)\rangle}
\end{align}
to denote quantum expectation.
It is straightforward to show that this definition satisfies (\ref{BMotion}). Note that it is the increments $dW_x$ which 
represent the physical stochastic information in our model. As stated above, for a given initial state and a complete set of realized values 
$\{dW_x\}$ the final state is uniquely specified by equation (\ref{SE}). The construction (\ref{Pnoise}) is just a useful 
way in which we can represent the stochastic 
information but we should be aware that the realized values $dB_x$ are not physical meaningful in the sense that they depend 
on the specific choice of foliation (via the state defined on surface $\sigma$). A different foliation would require a 
different realized $\mathbb{P}$-Brownian motion field to achieve the same evolved state on a given leaf. 

Expressing equation (\ref{SE}) directly in terms of the $\mathbb{P}$-Brownian motion field
we end up with the following nonlinear equation for the normalized state:
\begin{align}
\rd_x |\Psi(\sigma)\rangle = \Big\{ 
-i J(x) A(x) \rd \omega_x - & \half \lambda^2 \left[N(x)-\langle N(x)\rangle_{\sigma}\right]^2 \rd \omega_x  
\nonumber \\
&  + \lambda^{} \left[N(x)-\langle N(x)\rangle_{\sigma}\right]\rd B_x
\Big\}|\Psi(\sigma)\rangle,
\label{SEP}
\end{align}
with
$|\Psi(\sigma)\rangle = |\Phi(\sigma)\rangle \langle\Phi(\sigma)|\Phi(\sigma)\rangle^{-\half}$.
This equation enables us to generate physical sample paths for the state in terms of Brownian increments 
generated under the physical measure for a given spacetime foliation. By construction we know that this equation gives foliation independent 
results even though the nonlinearity obscures this fact.

\section{Local beables}
\label{lb}

If the model outlined above is to solve any of the conceptual problems of quantum theory then it must be equipped with a 
prescription for determining definite properties of the world in bounded regions of spacetime. Bell introduced the 
concept of local beables to provide such a means of describing a system in classical terms in order to make a 
clear point of contact with evidence of real world phenomena \cite{Bell}.

In describing state vector collapse our mathematical framework contains only the state vector and a classical
stochastic noise field. We first consider the former. Local properties of the state vector, 
as described by the action of local operators are indefinite for two reasons: (i) the state may not 
be an eigenstate of the operator in question; and (ii) nonlocalities ensure that the action of local operators 
on the state are affected by distant collapses which may or may not have happened depending on the choice of 
spacelike hypersurface. 

To address this Ghirardi \cite{Ghir4} has proposed that definite properties of the theory at point $x$ can be 
defined as the quantum expectations of local operators $O(x)$, where the state is assigned to the hypersurface $plc(x)$ 
forming the past light cone of $x$ (or the spacelike surface which is arbitrarily close to this):
\begin{align}
\bar{O}(x) = \langle O(x) \rangle_{plc(x)}.
\end{align}
Assuming that this past light cone limit is valid, we can define local beables in this way which are unambiguous, 
Lorentz covariant, and frame independent. This does not affect the arbitrary choice of foliation used to describe 
the state evolution---by conditioning on ${\cal F}_{\sigma}$ for any hypersurface $\sigma$ passing through the 
point $x$, the past light cone state is specified. 

It should not be necessary to grant beable status to the quantum expectation of every local operator in this way. Ghirardi
suggests that only the matter density need be a beable since this is enough to specify the 
locations of macro objects. In the spirit of relativity we suggest that the beables of the theory could be the stress-energy 
density of the quantum field
\begin{align}
\bar{T}^{\mu\nu}(x) = \langle T^{\mu\nu}(x) \rangle_{plc(x)},
\end{align}
(assuming that quantum expectations are finite following renormalization).

The other possible choice for the local beable of the theory is the classical stochastic noise field. From equation 
(\ref{Pnoise}) we can associate physical random variables to any finite region of spacetime $R$ as follows:
\begin{align}
W_R = \int_{R}dW_x = \int_{R}dB_x + 2\lambda \int_R d\omega_x \langle N(x) \rangle_{\sigma}.
\end{align}
This is a Lorentz invariant random variable (the explicit $\sigma$ dependence is offset by the foliation dependence of $dB_x$).
The right side of this equation demonstrates that the physical variable $W_R$ is composed of a signal---the quantum expectation of the 
operator $N(x)$ integrated over the region $R$---and a noise $B_{R} =\int_{R}dB_x$. For regions where the quantum expectation of $N(x)$ is large we 
can expect a large signal to noise ratio. The random variable $W_R$ then gives a classical image of $N(x)$-density.
This is perhaps the more natural choice for the beables given that the physical noise field is 
the classical element of the theory. However, note that the essential information in $W_R$ is (given in a Lorentz invariant form by)
$\bar{N}(x)$. This shows an equivalence between the two proposals. 

In each case, whether or not these variables are treated as local beables, they are nevertheless well defined Lorentz covariant
and frame independent local properties of the theory.

\section{Smeared operators}
\label{so}

Our dynamical equation for the state vector (\ref{SEP}) involves two different operators acting on the pointer 
field state:
\begin{align} 
N(x) = \int \rd\omega_y f(x,y) n(y)
\;\; ; \;\;
A(x) = \int \rd\omega_y g(x,y) \left[a(y)+a^{\dagger}(y)\right].
\nonumber
\end{align}
Each of these operators potentially describes a nonlocal interaction. 
The aim of this section is to consider some possible forms for the smearing functions $f$ and $g$ which satisfy the 
constraint of Lorentz invariance.

First consider the function $g(x,y)$. In order that the smeared interaction satisfies a reasonable definition of locality we would like 
for $g(x,y)$ to be appreciable only for points $y$ which are near to $x$. However, the notion of $y$ being near to $x$ 
is frame dependent. We therefore propose to use the local properties of the theory to determine the form of 
$g(x,y)$ in a way which takes account of the local energy flow of the field at point $x$. Specifically we propose a form 
\begin{align}
g(x,y)= 
C(x)\exp\left\{- k \bar{T}^{\mu\nu}(x)(y_{\mu}-x_{\mu})(y_{\nu}-x_{\nu})  \right\}, 
\label{g}
\end{align}  
for $y$ in the future cone of $x$ and $g(x,y)= 0$ elsewhere. Here $k$ is a positive real constant which controls the 
rate of decay of $g$ with $y$ and $C(x)$ is a positive real 
normalization function defined such that $g$ satisfies $\int d\omega_y g(x,y) = 1$. 
Note that the form of $g$ is Lorentz invariant. The exponent is negative definite since for $y$ within the future cone of $x$ we can 
always choose a frame in which $(y-x)$ defines the time direction and in this frame only the positive definite 
$\bar{T}^{00}$ component contributes. The stress-energy factor ensures that the function decays more rapidly in 
those timelike directions in which the magnitude of the field momentum is large (e.g. at the extremes of the light cone). 
The function $g(x,y)$ thus defines a distribution of points $y$ near to $x$ from the point of view of a field rest frame 
at $x$. Of course this function could take many other forms---this example is intended as an illustration.
 
Similarly for $f(x,y)$ we choose the smeared form 
\begin{align}
f(x,y)= 
C(x)\exp\left\{- k \bar{T}^{\mu\nu}(x)(x_{\mu}-y_{\mu})(x_{\nu}-y_{\nu})  \right\}, 
\label{FT}
\end{align}  
for $y$ in the past cone of $x$ and $f(x,y)=0$ elsewhere.

Since $\bar{T}^{\mu\nu}(x)$ is involved in the equations of motion of the state then the model is nonMarkovian: In order to 
advance the state from some arbitrary surface $\sigma$ to another surface $\sigma'$ which differs
from $\sigma$ only at the point $x$ we must determine $\bar{T}^{\mu\nu}(x)$. Since this depends of the surface $plc(x)$ we require
stochastic information encoded in $\{dW_y\}$ for $y$ to the past of $\sigma$ (but outside 
the past cone of $x$). This makes exact calculations difficult to perform.

\section{Collapse process}
\label{cp}

In this section we consider a specific example involving an initial superposition of different quantum matter density 
states. By decomposing into different time stages where either the field interactions dominate or the 
collapse process dominates we will demonstrate the characteristics of the state dynamics.
Let us express the initial state (assigned to an initial hypersurface $\sigma_i$) as a direct product of the quantum
field state (describing matter) and the pointer state
\begin{align}
|\Psi(\sigma_i)\rangle = |\Psi_{\rm matter}\rangle  |\Psi_{\rm pointer}\rangle.
\label{formINIT}
\end{align}
We contrive a situation in which the matter field is initially in a superposition of idealized $J(x)$-eigenstates, i.e.
\begin{align}
|\Psi_{\rm matter}\rangle  = \sum_i c_i |J_i\rangle,
\end{align}
where $|J_i\rangle$ are normalized and satisfy $J(x)|J_i\rangle = J_i(x)|J_i\rangle$
($J_i(x)$ is some real valued function of $x$). The initial condition of the pointer field (prior to interaction) is the 
ground state. Equation (\ref{formINIT}) can thus be written
\begin{align}
|\Psi(\sigma_i)\rangle =  \sum_i c_i |J_i\rangle  |0\rangle.
\label{init}
\end{align}
A state of this type could, for example, be formed following the interaction of some (quantum) measuring device with
a quantum particle (prior to any interaction with the pointer field).

If we ignore for now the collapse dynamics by setting $\lambda=0$ in equation (\ref{SEP}), we have the 
state evolution equation
\begin{align}
\rd_x |\Psi(\sigma)\rangle = -i J(x) A(x) \rd \omega_x |\Psi(\sigma)\rangle.
\end{align}
This equation has the formal solution
\begin{align}
|\Psi(\sigma)\rangle = \exp\left\{-i \int_{\sigma_i}^{\sigma} d\omega_x J(x) A(x) \right\} |\Psi(\sigma_i)\rangle,
\end{align}
where the terms in the exponent should be time ordered (noting that in general $[J(x),J(x')]\neq 0$ for timelike 
separated $x$ and $x'$).
Applying this solution to our initial condition (\ref{init}) we find that after the system has evolved to some
hypersurface $\sigma_{\rm int}$ (denoting the end of this pure interaction phase) the state is given by
\begin{align}
|\Psi(\sigma_{\rm int})\rangle = \sum_i c_i |J_i\rangle |\alpha_i\rangle,
\label{JAsol}
\end{align}
where
\begin{align}
|\alpha_i\rangle  = \exp\left\{\int d\omega_y \left[\alpha_i(y,\sigma_{\rm int})a^{\dagger}(y) 
- \alpha_i^*(y,\sigma_{\rm int})a(y)\right] \right\}|0\rangle,
\label{coh}
\end{align}
and
\begin{align}
\alpha_i(y,\sigma_{\rm int}) = -i  \int_{\sigma_i}^{\sigma_{\rm int}} d\omega_x  J_i(x)  g(x,y).
\label{coheig}
\end{align}
From equation (\ref{coh}) it is straightforward to show that the state $|\alpha_i\rangle$ has the property
\begin{align}
a(z)|\alpha_i\rangle = \alpha_i(z,\sigma_{\rm int})|\alpha_i\rangle.
\end{align}
The pointer field state is therefore a coherent state. 
Equation (\ref{coheig}) entails that the pointer field is excited in proportion to the matter density and is only excited 
in locations near where the matter density is nonzero (see equation (\ref{g})).
This analysis shows that a superposition state in the matter field leaves an imprint on the pointer field. 
An initial superposition of different $J(x)$-states results in an entangled superposition of $a$-eigenstates
after a short period of interaction. Notice that this will lead to a loss of coherence for the matter field state in cases where
the pointer field is significantly excited (the pointer field behaves as an environment) .

If the pointer field is in the state $|\alpha_i\rangle$, then the quantum expectation value of the operator $N(x)$ is
\begin{align}
\langle \alpha_i |N (x)|\alpha_i \rangle = \int d\omega_y f(x,y) |\alpha_i(y,\sigma_{\rm int})|^2,
\end{align}
and the quantum variance of the operator $N(x)$ is
\begin{align} 
\langle \alpha_i |N^2 (x)|\alpha_i \rangle - \langle \alpha_i |N (x)|\alpha_i \rangle^2 = 
\int d\omega_y f^2(x,y) |\alpha_i(y,\sigma_{\rm int})|^2.
\end{align}
This means that the size of quantum fluctuations in $N(x)$ behaves approximately as the square 
root of the expected value. Therefore, for sufficiently large values of $J_i(x)$ (corresponding to macroscopic matter) 
we can make the assumption that $|\alpha_i\rangle$ is an approximate $N(x)$-eigenstate:
\begin{align}
N(x)|\alpha_i\rangle \simeq \int d\omega_y f(x,y) |\alpha_i(y,\sigma_{\rm int})|^2|\alpha_i\rangle.
\label{Neig}
\end{align}

We now turn to the collapse dynamics, ignoring the $J(x)A(x)$ interaction term in equation (\ref{SEP}):
\begin{align}
\rd_x |\Psi(\sigma)\rangle = \Big\{ 
-  \half \lambda^2 & \left[N(x)-\langle N(x)\rangle_{\sigma}\right]^2 \rd \omega_x  
\nonumber \\ &  
+ \lambda^{} \left[N(x)-\langle N(x)\rangle_{\sigma}\right]\rd B_x
\Big\}|\Psi(\sigma)\rangle.
\label{collonly}
\end{align}
We take the state to be of the idealized postinteraction form 
\begin{align}
|\Psi(\sigma_{\rm int})\rangle = \sum_i c_i |J_i\rangle |N_i\rangle,
\label{perfectN}
\end{align}
where $|N_i\rangle$ satisfies $\langle N_i|N_j\rangle=\delta_{ij}$ and $N(x)|N_i\rangle = N_i(x)|N_i\rangle$
($N_i(x)$ is a real valued function of $x$). Denoting the quantum variance of the operator $N(x)$ as
\begin{align}
{\rm Var}_{\sigma}[N(x)] = \langle N^2(x) \rangle_{\sigma} - \langle N(x) \rangle^2_{\sigma},
\label{varperfectN}
\end{align}
we find that for a state of the form (\ref{perfectN}),
\begin{align}
{\rm Var}_{\sigma_{\rm int}}[N(x)] = \sum_i |c_i|^2 N^2_i(x) - \left( \sum_i |c_i|^2 N_i(x) \right)^2.
\end{align}
The quantum variance is greater than or equal to zero, and is only equal to zero if either (i) $|c_j|=1$ for some $j$ and 
$c_{i\neq j}=0$ for all other $i$s, or (ii) all $N_i(x)$s have the same value. We assume that the second situation is 
not true everywhere.

Similarly let us define the quantum covariance of $N(x)$ and $N(y)$ by
\begin{align}
{\rm Cov}_{\sigma}[N(x),N(y)] = \langle N(x)N(y) \rangle_{\sigma} - \langle N(x) \rangle_{\sigma}\langle N(y) \rangle_{\sigma}.
\label{cov}
\end{align} 

From equation (\ref{collonly}) we can show that the quantum variance of $N(x)$ satisfies the process
\begin{align}
\rd_y {\rm Var}_{\sigma}[N(x)] =  -4 & \lambda^2  {\rm Cov}^2_{\sigma}[N(x),N(y)] \rd \omega_y
\nonumber\\
 + 2\lambda & \left[  \langle N^2(x)N(y)\rangle_{\sigma} 
- \langle N^2(x)\rangle_{\sigma}\langle N(y)\rangle_{\sigma} \right. 
\nonumber \\ 
& \left. - 2\langle N(x)\rangle_{\sigma} \langle N(x)N(y)\rangle_{\sigma} 
+ 2\langle N(x)\rangle^2_{\sigma}\langle N(y)\rangle_{\sigma}\right]\rd B_y.
\end{align}
From this equation we find 
\begin{align}
\mathbb{E}^{\mathbb{P}}\left[\left. {\rm Var}_{\sigma}[N(x)] \right| {\cal F}_{\sigma_{\rm int}}\right] 
= & {\rm Var}_{\sigma_{\rm int}}[N(x)] \nonumber\\
 & - 4\lambda^2 \mathbb{E}^{\mathbb{P}}\left[\left. \int_{\sigma_{\rm int}}^{\sigma}\rd \omega_y{\rm Cov}^2_{\sigma'}[N(x),N(y)] 
\right| {\cal F}_{\sigma_{\rm int}}\right],
\label{varexp}
\end{align}
where the surfaces $\sigma'$ define a foliation between $\sigma_{\rm int}$ and $\sigma$ and $y\in \sigma'$. 
In general for nonzero covariance of $N$, equation 
(\ref{varexp}) indicates that the $\mathbb{P}$-expected quantum variance will decrease as $\sigma$ advances through spacetime. 
Since the $\mathbb{P}$-expectation of quantum variance tends to zero then the realized quantum variance must tend to zero. This 
implies that the state tends to an $N(x)$-eigenstate on a collapse timescale of order
\begin{align}
\tau_{\rm coll} \sim \frac{ {\rm Var}_{\sigma_{\rm int}}[N(x)]}{\lambda^2 \int d^3 y {\rm Cov}^2_{\sigma_{\rm int}}[N(x),N(y)]}
\label{tau}
\end{align}
(in the frame defined by our chosen time slice).

Now consider the projection operator $P_j = |N_j\rangle \langle N_j|$. Given equation (\ref{collonly}), the quantum expectation of $P_j$
satisfies 
\begin{align}
d_x \langle P_j \rangle_{\sigma} = \lambda \langle \{ P_j, N(x)\}\rangle_{\sigma}dB_x
-2\lambda \langle  P_j\rangle_{\sigma}\langle N(x) \rangle_{\sigma}dB_x.
\end{align}
This means that $\langle P_j \rangle_{\sigma}$ is a $\mathbb{P}$-martingale, i.e.
\begin{align}
\mathbb{E}^{\mathbb{P}}\left[\left. \langle P_j \rangle_{\sigma} \right| {\cal F}_{\sigma_{\rm int}}\right]
=\langle P_j \rangle_{\sigma_{\rm int}},
\end{align}
for $\sigma_{\rm int}\prec \sigma$. As the quantum variance of $N(x)$ tends to zero then either $\langle  P_j\rangle_{\sigma}\rightarrow 1$ or 
$\langle  P_j\rangle_{\sigma}\rightarrow 0$ depending on whether the state ends up as $|N_j\rangle$ or not. 
Let $\sigma_{\rm coll}$ denote the end of this collapse phase where we can apply these limits. 
We have
\begin{align}
\mathbb{E}^{\mathbb{P}}\left[\left. \langle P_j \rangle_{\sigma_{\rm coll}} \right| {\cal F}_{\sigma_{\rm int}}\right]
=\mathbb{E}^{\mathbb{P}}\left[\left. 1_{\{ |\Psi(\sigma_{\rm coll})\rangle = |N_j\rangle \}} \right| {\cal F}_{\sigma_{\rm int}}\right]
=\langle P_j \rangle_{\sigma_{\rm int}}. 
\end{align}
This equation states that the stochastic probability of a given outcome (in this case $|\Psi(\sigma_{\rm coll})\rangle = |N_j\rangle$)
is given by the initial quantum prediction for the probability of this outcome ($\langle P_j \rangle_{\sigma_{\rm int}}$),
i.e. the Born rule is satisfied.

For the initial state given by (\ref{init}) the result is therefore
\begin{align}
|\Psi(\sigma_{\rm coll})\rangle \simeq
|J_i\rangle |\alpha_i\rangle & \text{ with prob } |c_i|^2.
\label{colstate}
\end{align}

Consider an equal superposition ($|c_1|=|c_2|$) of two matter field states with eigenvalues $J_i(x)$ such that,
in the rest frame of the system, $J_1(x)=J$ only in the spatial region $R_1$ ($J_1(x)=0$ elsewhere) and $J_2(x)=J$ only 
in the spatial region $R_2$. Further consider $V_{\triangle}$ to be the spatial volume of the symmetric difference $R_1\triangle R_2$
(containing points belonging to one but not both of $R_1$ and $R_2$). 
Using equations (\ref{tau}), (\ref{cov}), (\ref{varperfectN}), (\ref{Neig}), and (\ref{coheig}),
and assuming that the smearing scales associated with $f(x,y)$ and $g(x,y)$ are much smaller 
than the scale associated with $V_{\triangle}$, we find $\tau_{\rm coll}\sim \lambda^{-2}V_{\triangle}^{-1} J^{-4}$. 
The rate of reduction depends on the magnitude of the matter density eigenvalue $J$ and the spatial extent 
of the region $R_1\triangle R_2$. This amplification effect ensures that low energy excitations can be essentially 
unaffected by the collapse mechanism whilst large scale superpositions (as characterized by $J$ and 
$V_{\triangle}$) undergo rapid state reduction. The precise rate is controlled by the stochastic coupling parameter.

So far we have considered the dynamics of the state vector. In order to describe the system in definite terms
we must consider the dynamics of the local properties of the theory. 
In order to estimate the stress-energy density we make the simplifying assumption that in the rest frame of the 
matter field, only the $T^{00}$ component is nonzero. This corresponds to the assumption that the matter behaves
as a swarm of noninteracting particles.  In this frame we assume that 
\begin{align}
T^{00}(x) |J_i\rangle = 
E_i(x)|J_i\rangle,
\end{align}
with all other components equal to zero. We further assume $E_i(x)$ and $J_i(x)$ are related in that they agree with
regards to the approximate distribution of matter. With the state given by equation (\ref{JAsol}) 
(with $\sigma_{\rm int}=plc(x)$), the stress energy density beable in the matter field rest frame is
\begin{align}
\bar{T}^{00}(x) = \sum_i |c_1|^2 E_i (x).
\end{align}
After collapse when the state takes the form of equation (\ref{colstate}) 
(with $\sigma_{\rm coll}=plc(x)$), $\bar{T}^{00}(x)$ is equal to $E_i(x)$ with probability $|c_i|^2$.

\section{Energy process}
\label{ep}

We have seen in previous sections that the collapse terms in the dynamical equation for the state vector involve 
the operators $N(x)$ and that this results in collapse toward an $N(x)$-eigenstate. If we had chosen, for example, 
a scalar field operator $\varphi(x)$ in place of $N(x)$ we would expect collapse toward a $\varphi(x)$-eigenstate. 
The problem in this case is that, as the $\varphi(x)$-state becomes more certain, the scalar field momentum state 
becomes more uncertain. The result is a divergent increase in the energy density \cite{pear3, pearGhir}. 

Here we demonstrate that the present model does not suffer from this problem. A sensible choice for the energy of the
pointer field is given by
\begin{align}
H_{\rm pointer} = \int d \omega_x a^{\dagger}(x) i \partial_{x_0} a(x).
\end{align}
This operator generates time translations in the pointer field annihilation and creation operators:
\begin{align}
[H_{\rm pointer},a(x)] = -i\partial_{x_0} a(x)
\;\; ; \;\;
[H_{\rm pointer},a^{\dagger}(x)] = -i\partial_{x_0} a^{\dagger}(x).
\end{align}
For the operator $A(x)$ we have
\begin{align}
[H_{\rm pointer},A(x)] = -\int d\omega_y g(x,y) i\partial_{y_0}\left[ a(y) + a^{\dagger}(y) \right].
\end{align}
The pointer field energy does not generate time translations in $A(x)$ unless we make the assumption that 
$\partial_{x_0} g(x,y) \simeq -\partial_{y_0} g(x,y)$ valid when $\bar{T}^{\mu\nu}(x)$ is slowly varying with 
time (when compared to $g(x,y)$). We can then integrate by parts to find
\begin{align}
[H_{\rm pointer},A(x)]\simeq -i\partial_{x_0} A(x).
\label{AComAp}
\end{align}
With this approximation the $J(x)A(x)$ interaction term will conserve a total energy of the form
\begin{align} 
H_{\rm total} = H_{\rm matter} + H_{\rm pointer} + \int d^3 x J(x) A(x),
\end{align}
provided that $H_{\rm matter}$ satisfies $[H_{\rm matter},J(x)] = -i\partial_{x_0} J(x)$.

We expect the collapse terms in the equations of motion to result in nonconservation of energy since the collapse 
process should be able to randomly choose from a superposition of differing energy states. Given some 
operator ${O}$ we can show using equation (\ref{SEP}) that its quantum expectation satisfies
\begin{align}
\rd_x\langle {O} \rangle_{\sigma} = {}& 
\langle\rd_x{O}\rangle_{\sigma} 
-i  \langle\left[{O},  J(x)A(x)\right]\rangle_{\sigma} \rd\omega_x
\nonumber\\
& -\half\lambda^2\langle\left[N(x),\left[N(x),{O}\right]\right]\rangle_{\sigma}\rd\omega_x \nonumber\\
& +\lambda\langle\left\{{O},N(x)\right\}\rangle_{\sigma}\rd B_x 
- 2\lambda\langle{O}\rangle_{\sigma} \langle N(x)\rangle_{\sigma}\rd B_x.
\label{opexp}
\end{align}
For example, setting $O=H_{\rm matter}$ we find
\begin{align}
\rd_x\langle {H_{\rm matter}} \rangle_{\sigma} = & -  \langle A(x) \partial_{x_0} J(x) \rangle_{\sigma} \rd\omega_x
\nonumber\\
& +\lambda\langle\left\{{H_{\rm matter}},N(x)\right\}\rangle_{\sigma}\rd B_x 
- 2\lambda\langle{H_{\rm matter}}\rangle_{\sigma} \langle N(x)\rangle_{\sigma}\rd B_x.
\label{mattereng}
\end{align}
Any changes in energy described by equation (\ref{mattereng}) must be consistent with experimental bounds 
on energy conservation. This will result in bounds on the parameters of the model.

For the pointer field energy,
\begin{align}
\rd_x\langle {H_{\rm pointer}} \rangle_{\sigma} \simeq {}& 
-  \langle J(x)  \partial_{x_0} A(x)  \rangle_{\sigma} \rd\omega_x
\nonumber\\
& -\half\lambda^2\langle\left[N(x),\left[N(x),{H_{\rm pointer}}\right]\right]\rangle_{\sigma}\rd\omega_x \nonumber\\
& +\lambda\langle\left\{{H_{\rm pointer}},N(x)\right\}\rangle_{\sigma}\rd B_x 
- 2\lambda\langle{H_{\rm pointer}}\rangle_{\sigma} \langle N(x)\rangle_{\sigma}\rd B_x.
\label{pointereng}
\end{align}
where we have made use of equation (\ref{AComAp}). To calculate the second term on the right side we use
\begin{align}
[N(x),H_{\rm pointer}] = \int d \omega_y f(x,y) \left[\left(i\partial_{x_0}a^{\dagger}(x)\right) a(x) 
+  a^{\dagger}(x) \left(i\partial_{x_0}a(x)\right)\right],
\end{align}
which in turn can be used to show that
\begin{align}
\left[N(x),\left[N(x),{H_{\rm pointer}}\right]\right] = 0.
\end{align}
There are no divergences in the rates of change of either $H_{\rm matter}$ or $H_{\rm pointer}$. There is
exchange of energy between the two fields driven by the $J(x)A(x)$ interaction but the collapse process 
conserves total energy in expectation.

Let us briefly consider how this model is affected by letting the smearing function $f$ become a delta function. 
In this case we have $N(x)=n(x)$ and we find
\begin{align}
\left[n(x),\left[n(x),{H_{\rm pointer}}\right]\right] = -\left(i\partial_{x_0}a^{\dagger}(x)\right) a(x) \delta^4(0)
 +  a^{\dagger}(x) \left(i\partial_{x_0}a(x)\right) \delta^4(0).
\end{align}
The pointer field energy therefore changes at an infinite rate due to the collapse dynamics. It can also be 
shown that taking $g(x,y)=\delta^4(x-y)$ leads to divergences for the matter field energy.

\section{Numerical calculation in 2D}

In order to understand the collapse dynamics in more detail we consider a numerical solution of the state evolution
in the simplified case of a 2D spacetime (one space and one time dimension). We use the example of an initial 
superposition of two different matter density states. 

Consider the state $|J_i\rangle |\alpha_i\rangle$ where in the rest frame of the 
matter field we have
\begin{align}
J({x_1}, x_0)|J_i\rangle = 
\begin{cases}
J_i|J_i\rangle & \text{for } {l}_i < {x_1} <{u}_i,\\
0 & \text{otherwise},
\end{cases}
\end{align}
where $l_i$ and $u_i$ are constant lower and upper bounds respectively of the spatial extent of the matter density. 
Assuming a sufficiently large value for the eigenvalue $J_i$ we can make the approximation (see section \ref{cp})
\begin{align}
N(x_1, x_0)|\alpha_i\rangle \simeq N_i (x_1,x_0)|\alpha_i\rangle.
\end{align}
We further assume that the length scale associated with $f$ and $g$ is sufficiently small (compared to length scales 
$u_i-l_i$ and their overlaps) that we can approximate equations (\ref{coheig}) and (\ref{Neig}) to give
\begin{align}
N_i({x_1}, x_0) = 
\begin{cases}
J^2_i & \text{for } {l}_i < {x_1} <{u}_i,\\
0 & \text{otherwise}.
\end{cases}
\label{Nsol}
\end{align}
The pointer field thus provides an image of the matter field state.

Suppose that the state of the system following some interaction between matter field and pointer field is
\begin{align}
|\Psi(\sigma)\rangle = c_1 |J_1\rangle |\alpha_1\rangle  + c_2 |J_2\rangle |\alpha_2\rangle.
\end{align}
In the rest frame we consider matter density states which are nonzero 
only in the following regions: $l_1 = -1 ; u_1 =0$ and $l_2 = 0 ; u_2 =1$. This corresponds to an initial 
superposition of two adjacent lumps of matter.

\begin{figure}[t]
\begin{center}
\includegraphics[width=10cm]{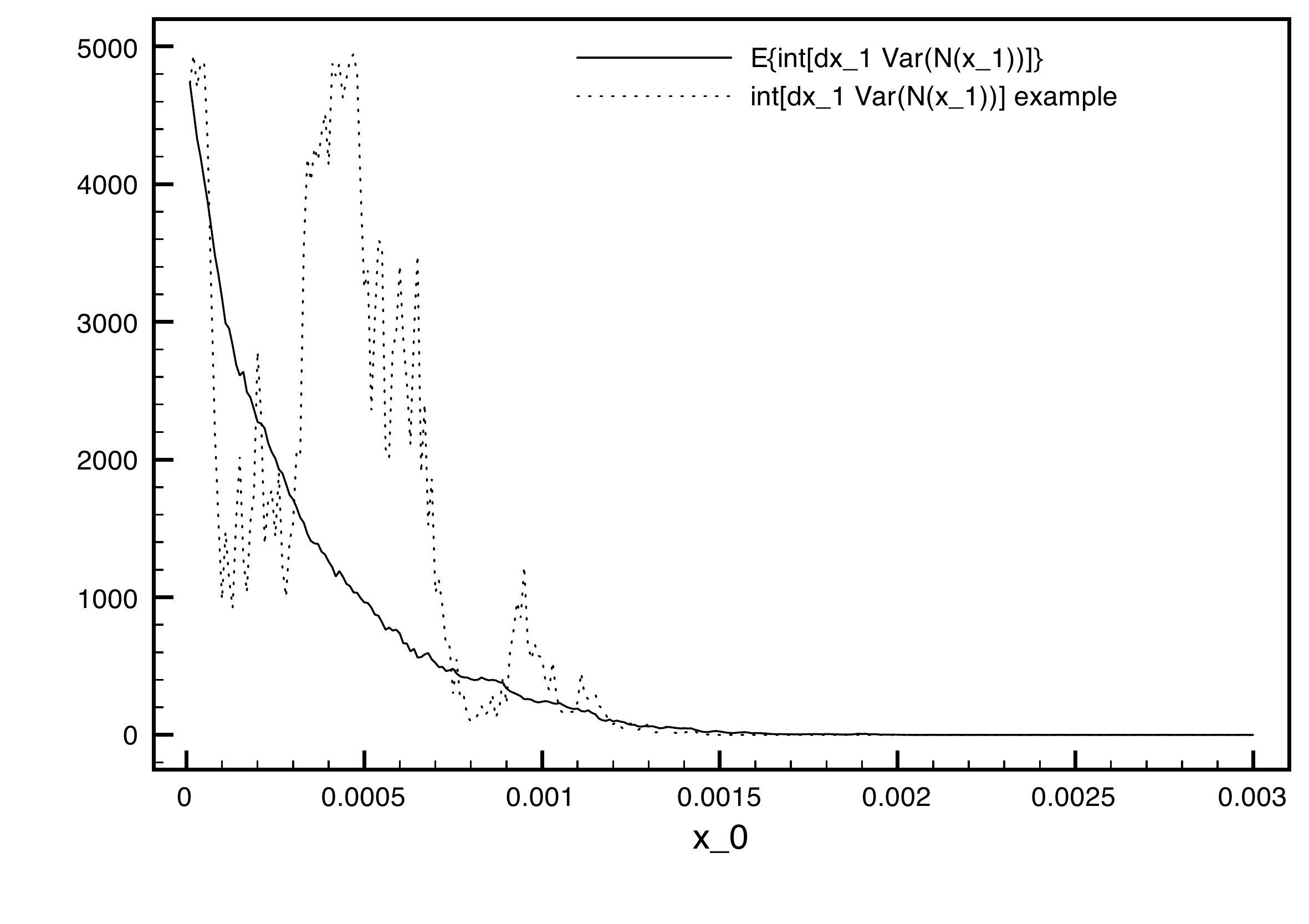}
\caption{
{
Numerical demonstration of the collapse process. The variable 
$\mathbb{E}^{\mathbb{P}}\left[ \int dx_1 {\rm Var}_{\sigma}[N(x_1,x_0)]\right]$ tends to zero as the system evolves
indicating that only one of the $N$-eigenstates survives. An example path highlights the stochastic nature of a typical
realized process.
}}
\end{center}
\end{figure}

We can solve equation (\ref{SEP}) numerically for this two-state example. Since the $|\alpha_i\rangle$ 
are approximate eigenstates of the operator $N(x_1, x_0)$, 
the state dynamics reduces to the dynamics of the two coefficients $c_1$ 
and $c_2$ (taken to be initially equal in our simulation). For simplicity the state evolution is considered only in terms of a foliation of constant $x_0$ surfaces. 
The stochastic coupling $\lambda$ is set to be equal to 0.5 and $J_i^2$ is 100 for both $i=1,2$.

In order to characterize the collapse process we use the quantity
\begin{align}
\int dx_1 {\rm Var}_{\sigma}[N(x_1,x_0)].
\label{collmeas}
\end{align}
As this quantity tends to zero then the state must tend to a $N(x_1,x_0)$-eigenstate for all $x_1$. Figure 2 shows how the 
$\mathbb{P}$-expectation of (\ref{collmeas}) decreases as the system evolves. The $\mathbb{P}$-expectation is calculated by 
Monte Carlo simulation using 200 sample paths. An example path is shown in the figure to highlight the stochastic nature
of the realized process.

Since the $\mathbb{P}$-expectation of (\ref{collmeas}) tends to zero then the realized quantum variance of $N$ must with certainty tend to 
zero. The state ends up in the form $|\Psi(\sigma)\rangle = |J_i\rangle |\alpha_i\rangle$. With initial conditions 
specified by $c_1 = c_2 = 1/\sqrt{2}$ we find that the proportion of occurrences of 
$|\Psi(\sigma)\rangle \rightarrow |J_1\rangle |\alpha_1\rangle$ and 
$|\Psi(\sigma)\rangle \rightarrow |J_2\rangle |\alpha_2\rangle$ are even to within statistical error.
The timescale for collapse is of order $10^{-4}-10^{-3}$ (in units defined by the chosen parameters). This is well approximated by the
formula $\tau_{\rm coll}\sim \lambda^{-2}V_{\triangle}^{-1} J^{-4}$.

\section{Discussion}

In this article we have outlined a framework for describing the evolution of relativistic quantum systems 
which consistently explains the behavior of both microscopic and macroscopic systems. To do this the model incorporates 
quantum state reduction into the standard state dynamics in a way which is not only covariant and frame 
independent, but also objective, naturally differentiating between systems of different scale and adjusting 
its effect accordingly. In this way the model offers a potential unification of quantum and classical sectors.

Within this framework no judgment is required on when to apply collapse and when to apply unitary evolution (as with 
orthodox quantum theory) and it is not necessary to perform an arbitrary separation of system and environment in order 
to understand its decoherence properties. The present model leads to the prediction of well defined observer independent 
local properties.

The mechanism can be used to describe collapse in any quantum field for which we can form a Lorentz invariant scalar 
current $J(x)$. This applies to both fermions and bosons. There is no incompatibility with the inclusion of
gauge field interactions and there are therefore no problems in principle with application to the 
standard model of particle physics. (We note that the model of Tumulka \cite{Tum}, although of 
interest as a demonstration of a consistent and formally rigorous relativistic collapse model, applies only to a fixed number of 
noninteracting quantum particles. To consider any interesting correlations between particle states
in this model they must be encoded in the initial state vector.)

The present model has features with the potential for experimental scrutiny. For example, the expected rates of collapse show a 
dependence on the specific system details; local properties of the theory exhibit a well defined stochasticity; and
there is energy transfer between the quantum fields and the pointer field. By quantifying these effects it is hoped that 
new tests of quantum theory may be suggested.

\vspace{10pt}
\noindent
{\it Acknowledgments}. It is a pleasure to thank Philip Pearle, GianCarlo Ghirardi, Shelly Goldstein, and Fay Dowker 
for helpful discussions.

\end{document}